\documentclass[sigconf]{acmart}

\AtBeginDocument{%
  }

\setcopyright{acmlicensed}
\copyrightyear{2026}
\acmYear{2026}

\acmConference[WCCCE '26]{Western Canada Conference on Computing Education}{April 30--May 1, 2026}{Vancouver, Canada}
\acmBooktitle{Western Canada Conference on Computing Education (WCCCE '26), April 30--May 1, 2026, Vancouver, Canada}

\usepackage{graphicx}
\usepackage{xcolor}
\usepackage{subcaption}
\usepackage{enumitem}
\usepackage{booktabs}
\usepackage{array}

\begin{document}

\title{Large-Scale Analysis of Discussions by CS Educators Across the Stack Exchange Network}

\author{Farhad Hossain}
\affiliation{
  \institution{Trent University}
  \city{Peterborough}
  \state{Ontario}
  \country{Canada}
}
\email{mdfarhadhossain@trentu.ca}

\author{Natasha Grech}
\affiliation{
  \institution{Trent University}
  \city{Peterborough}
  \state{Ontario}
  \country{Canada}
}
\email{natashagrech@trentu.ca}

\author{Omar Alam}
\affiliation{
  \institution{Trent University}
  \city{Peterborough}
  \state{Ontario}
  \country{Canada}
}
\email{omaralam@trentu.ca}

% \author{\IEEEauthorblockN{ Md Farhad Hossain} % 1\textsuperscript{st}
% \IEEEauthorblockA{\textit{Trent University} \\
% \textit{Peterborough, ON}\\
%  Canada   \\
% \textit{mdfarhadhossain@trentu.ca}}

% \and

% \IEEEauthorblockN{ Omar Alam} % 1\textsuperscript{st}
% \IEEEauthorblockA{\textit{Trent University} \\
% \textit{Peterborough, ON}\\
%  Canada   \\
% \textit{omaralam@trentu.ca}}
 
% }

\begin{abstract}

Stack Exchange is a widely used question-and-answer network that facilitates knowledge exchange across diverse fields. Within this network, the Computer Science (CS) Educators Stack Exchange (CS Educators) serves as a focused space where CS educators share insights, ask questions, and discuss teaching practices. In this study, we examined 79,854,463 posts—comprising 32,187,805 questions and 47,666,658 answers—across the Stack Exchange network, with an emphasis on English-language posts contributed by CS Educators participants. Using topic modeling, we identified and manually labeled key topics, organized them hierarchically, and analyzed their distribution and complexity. Our analysis reveals evolving discussion patterns that reflect both technical (IT) and non-technical (Non-IT) themes. Within IT, topics related to programming and software development were most prominent, while mathematics, humanities, and education attracted significant attention within the Non-IT categories. These findings highlight the breadth of participation among CS Educators and shed light on their diverse priorities. We hope this work will help improve understanding of CS educators’ topical interests and support the development of strategies to better address and focus on these areas.

\end{abstract}




\maketitle

\section{Introduction}
%CS education is evolving rapidly due to technological change and the need for updated teaching practices. As tools and methods advance, educators must keep pace to remain effective.

Computer science education has experienced enormous interest in recent years due to the rapid growth of the technology sector. CS educators must keep pace to remain relevant in this rapidly advancing field. Online platforms on the Stack Exchange network \cite{stackexchangesites2024} have become important discussion spaces, offering peer-driven insights into the issues that developers and technology practitioners face. Major sites like Stack Overflow (25 million users, 24 million questions) \cite{stackoverflow2024}, Super User (1.7 million users) \cite{superuser2024}, and Ask Ubuntu (1.6 million users) \cite{askubuntu2024} have become a go-to place for practitioners seeking solutions to both technical and non technical issues.

The CS Educators site \cite{cseducators2024}, a smaller site within Stack Exchange with about 12,949 users, serves as a dedicated space for discussing pedagogy, course design, and instructional tools. These users discuss topics within this site but also across other sites within the broader Stack Exchange network.

%The range of viewpoints across Stack Exchange supports educators’ growth and helps them translate theory into practice—Stack Overflow alone features over 36 million answers, reflecting diverse problem-solving strategies. These platforms help educators adapt to shifting technological demands. This study investigates how CS educators use these online spaces to navigate instructional trends and emerging challenges.

In this paper, our aim is to analyze the contributions and discussions of CS Educators across Stack Exchange platforms. Our motivation is to understand what issues they are discussing, not only on the CS Educators site but also across Stack Exchange platforms, through the following research questions:

\textbf{RQ1: What topics are discussed by CS educators across all sites of Stack Exchange network?}

\textbf{RQ2: How do the topics discussed by CS educators evolve over time?}
%By tracking the evolution of these discussions, we can identify shifts in educational focus and emerging challenges.

% To address these questions, we analyzed posts from 169 English-language Stack Exchange sites, narrowing our scope to 748,084 entries authored by CS educators, including 397,061 questions and 351,023 accepted answers. Using topic modeling techniques \cite{blei2003}, we identified discussion topics, categorized them and studied how they evolve over time. In particular, we idneitified two broad cateogories, IT and Non-IT topics. IN IT, discussions on Programming and Software development tools dominated with topics of both subcategory of 34\% of dicussions. Among non-IT, Topics on Mathematics dominated with 8.22\% of discussions.
% To address these questions, we analyzed posts from 169 English-language Stack Exchange sites, narrowing our scope to 748,084 entries authored by CS educators, including 397,061 questions and 351,023 accepted answers. Using topic modeling techniques~\cite{blei2003}, we identified discussion topics, categorized them, and examined their evolution over time. In particular, we identified two broad categories: IT (64.88\%) and Non-IT (35.12\%) topics. Within the IT category, discussions on programming and software development tools dominated, together accounting for 34\% of the discussions. Among Non-IT topics, mathematics-related discussions were the most prominent, comprising 8.22\% of the discussions.

To address these questions, we analyzed posts from 169 English-language Stack Exchange sites, narrowing our scope to 748,084 entries authored by CS educators, including 397,061 questions and 351,023 accepted answers. Using topic modeling techniques~\cite{blei2003}, we identified discussion topics, categorized them, and examined their evolution over time. In particular, we identified two broad categories: IT (64.88\%) and Non-IT (35.12\%) topics. Within the IT category, discussions on programming and software development tools accounted for 34\% of the discussions. Among Non-IT topics, mathematics-related discussions were the most prominent, comprising 8.22\% of the discussions, followed by lifestyle-related discussions. Although IT-related topics dominate discussions among CS educators—especially those related to programming and software development—Non-IT categories experienced steady growth from 2013 to 2018.

We hope that our findings provide insight into the discussions and concerns of CS educators and help clarify general discussion trends. Policymakers, researchers, and educational technology developers can use these findings to better design solutions that support CS educators. Furthermore, CS educators and educational institutions may become more aware of the topics that attract the greatest interest among their peers and, accordingly, deploy targeted training resources and adjust policies to better address these areas.

The remainder of the paper is structured as follows: The next section discusses the related work. Section 3 outlines the methodology and data collection process. Section 4 answers RQ1 and Section 5 answers RQ2. Section~6 discusses the threats to validity. Section~7 concludes the paper and outlines potential future work.

\section{Related Work}

%As the demand for technically skilled professionals grows, computer science education is becoming increasingly vital. Platforms like CS Educators Stack Exchange provide a focused environment for instructors to share insights on pedagogy, curriculum, and classroom issues \cite{cseducators2024}. 

While previous studies have examined technical discussions on sites like Stack Overflow, they often focus on a single platform \cite{barua2012, bagherzadeh2019}. For instance, Barua et al. explored trends in software topics \cite{barua2012}, and Bagherzadeh and Khatchadourian addressed big data challenges \cite{bagherzadeh2019}. However, Stack Overflow is only one of 183 English-language sites within the Stack Exchange network, covering both technical and non-technical areas \cite{stackexchangesites2024}. Topic modeling has been used across this network to analyze a variety of domains. Tahir et al. examined code smells across platforms such as Stack Overflow and Code Review \cite{tahir2020}, while others explored continuous engineering \cite{zahedi2020}, microservices \cite{bandeira2019}, Text-to-SQL \cite{hazoom2021}, software privacy \cite{tahaei2020}, and new languages like Swift and Rust \cite{chakraborty2021}. Additional studies addressed technical debt \cite{santos2023}, Java \cite{blanco2020}, refactoring \cite{peruma2021}, IoT \cite{uddin2021}, chatbots \cite{abdellatif2020}, and Flutter \cite{alanazi2024}, yet few included educator-specific data. 

This study applies Latent Dirichlet Allocation (LDA) \cite{blei2003} to identify themes in discussions by CS educators across the entire Stack Exchange network. LDA has proven effective in analyzing large text corpora from Q\&A platforms, repositories, and social media \cite{sun2015}. For example, Chen et al. studied software logging, while Li focused on software maintenance \cite{chen2016}. Our approach captures how CS educators engage in both technical and broader discussions—such as law, history, and the arts—addressing a gap left by earlier work that focused mostly on technical forums. There has been some prior work focusing on the CS Eucators~\cite{omaralam2020, lal2022cs}. For example, Lal et al.~\cite{lal2022cs} studied issues related to online teaching by analyzing posts tagged with “online” and “distance learning” on CS Educators. Moudgalya et al.~\cite{moudgalya2019computer} conducted a qualitative study on CS Educators to understand perceptions of gender equity and diversity in computer science. However, to the best of our knowledge, no prior work has conducted an extensive analysis of discussions across the Stack Exchange network users of CS Educators, as we do in this paper.

\section{Data Collection and Methodology}
\subsection{Data Collection}
\begin{figure}[t]
    \centering
    \includegraphics[width=1\linewidth]{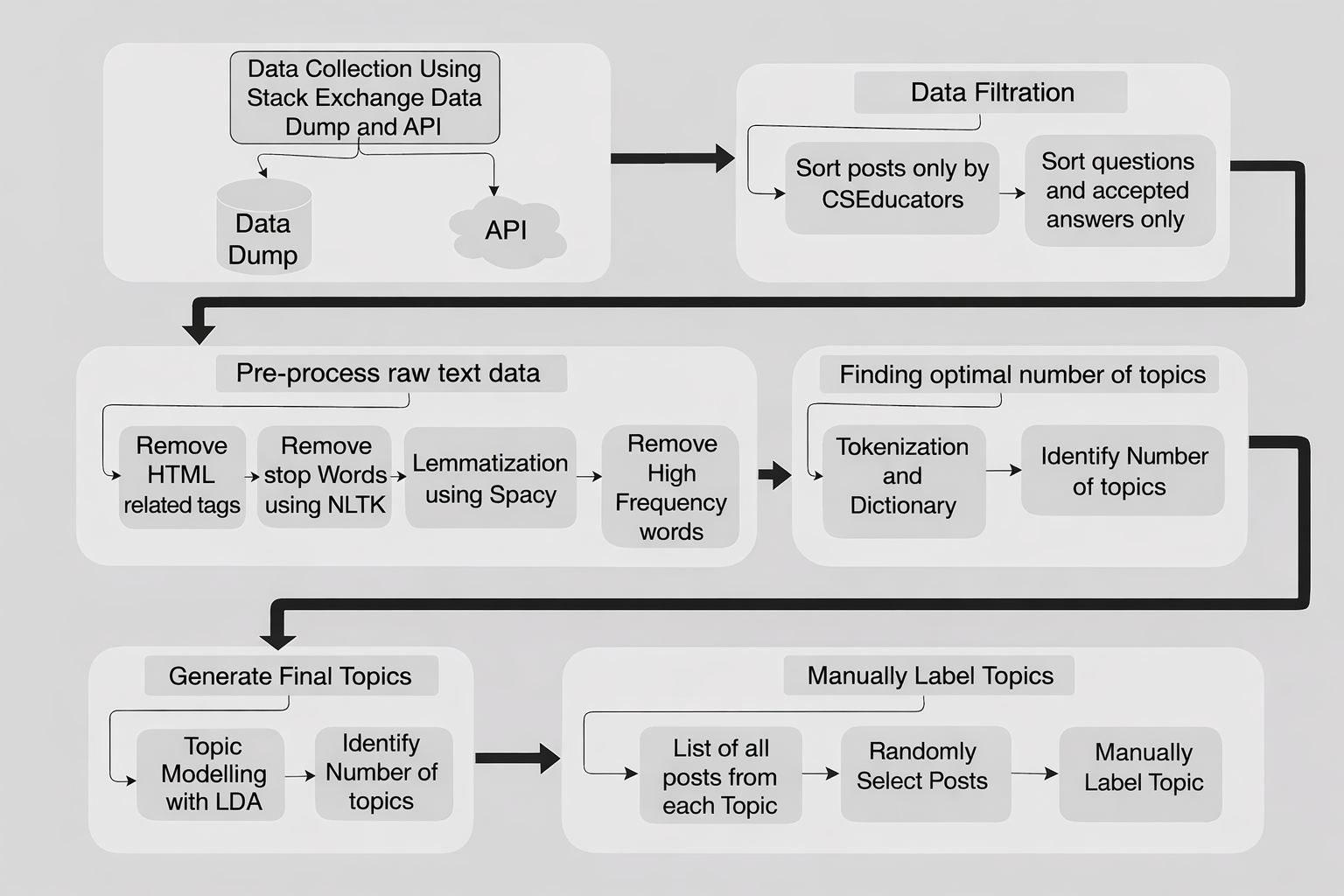}
    \caption{Overview Flow Chart}
    \label{fig:Topic Modeling Overview Flow Chart}
\end{figure}

\begin{figure}[t]
    \centering
    \includegraphics[width=1\linewidth]{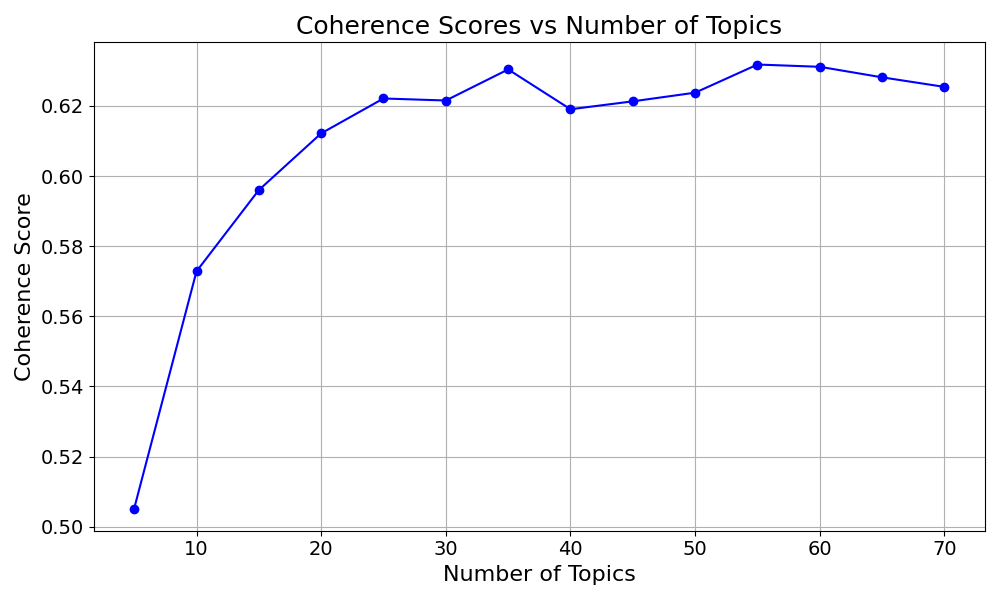}
    \caption{Topics-Coherence Scores Graph}
    \label{fig:Topics-Coherence score graph}
\end{figure}

We applied topic modeling across all questions, answers contributed by users of CS Educators throughout the entire Stack Exchange network, rather than limiting the analysis to a single site such as Stack Overflow. The network currently consists of 183 sites spanning a wide range of subjects, including both technical fields like programming and software engineering, as well as non-technical areas such as law and history. 

To gather the necessary data, we utilized two primary methods: 

\textbf{API} : The Stack Exchange API enables targeted, real-time retrieval of posts and user activity across sites \cite{stackexchangeapi2024}. 

\textbf{Data Dump}: Stack Exchange also releases a comprehensive data dump, which includes structured XML files containing site-wide data such as posts, users, tags, and votes \cite{stackexchangedatadump2024}. 

Each method has its trade-offs. While the API supports more specific queries, it is less efficient for large-scale data extraction. In contrast, the data dump provides a complete snapshot of all 183 sites, approximately 120 gigabytes in total enabling broader analysis at the expense of flexibility. Each site folder includes standardized XML files such as \texttt{Posts.xml}, \texttt{Users.xml}, \texttt{Tags.xml}, and \texttt{Votes.xml}. 
Although the Stack Exchange data dump provides a comprehensive snapshot of all 183 sites—totaling approximately 120 GB and including key XML files such as \texttt{Posts.xml}, \texttt{Tags.xml}, \texttt{Users.xml}, and \texttt{Votes.xml}—our analysis focused exclusively on the 169 sites that operate in English. User identifiers like \texttt{UserId} are unique only within individual sites and not across the entire network. In contrast, the \texttt{AccountId} field uniquely identifies users across all Stack Exchange platforms. However, a major drawback is that \texttt{AccountId} is not included in the \texttt{Posts.xml} file, making it difficult to trace a user's contributions across different sites using the data dump alone.

\begin{table}[t]
    \centering
    \caption{Summary of Stack Exchange Dataset}
    \label{tab:summary_posts}
    \begin{tabular}{@{}ll@{}}
        \toprule
        \textbf{Description}                         & \textbf{Value}       \\ \midrule
        Total number of posts on Stack Exchange      & 79,854,463           \\
        Total number of questions on Stack Exchange  & 32,187,805    \\
        Total number of answers on Stack Exchange    & 47,666,658           \\
        Total number of CS educators                 & 12,949               \\
        Total questions by CS educators              & 397,061              \\
        Total accepted answers                       & 351,023              \\
        Posts used for topic modeling                & 784,048              \\
        Dataset start date                           & 2008-03-02           \\
        Dataset end date                             & 2024-05-16           \\
        Total number of Stack Exchange sites                             & 169          \\
        \bottomrule
    \end{tabular}
\end{table}

To overcome this, we adopted a hybrid approach that leveraged both the data dump and the Stack Exchange API. While the data dump provided the majority of post content, the API was employed to retrieve cross-site user identifiers. In particular, we used the API endpoint  \url{https://api.stackexchange.com/2.3/users/{account\_ids}}. Then, for each question, we selected only the accepted answers and used them for further analysis.

After combining data from both the API and the data dump, we assembled a unified dataset containing 79,854,463 posts (32,187,805 questions and 47,666,658 answers) as of May 16, 2024 \cite{stackexchangesites2024}. After filtering for posts authored by CS educators and selecting only those with accepted answers, the final dataset comprised 748,084 entries, including 397,061 questions and 351,023 accepted answers as shown in Table~\ref{tab:summary_posts}. In line with previous studies~\cite{bagherzadeh2019, barua2012}, we analyzed only accepted answers (i.e. we excluded unaccepted answers) in order to reduce noise and manage the size of the dataset.

The final dataset compiles posts from the full Stack Exchange network, specifically filtered for CS Educators, offering a broad perspective on their engagement across both technical and non-technical domains.

\begin{figure*}[t]
     \centering
    \resizebox{.78\linewidth}{!}{
     \includegraphics[width=.5\textwidth]{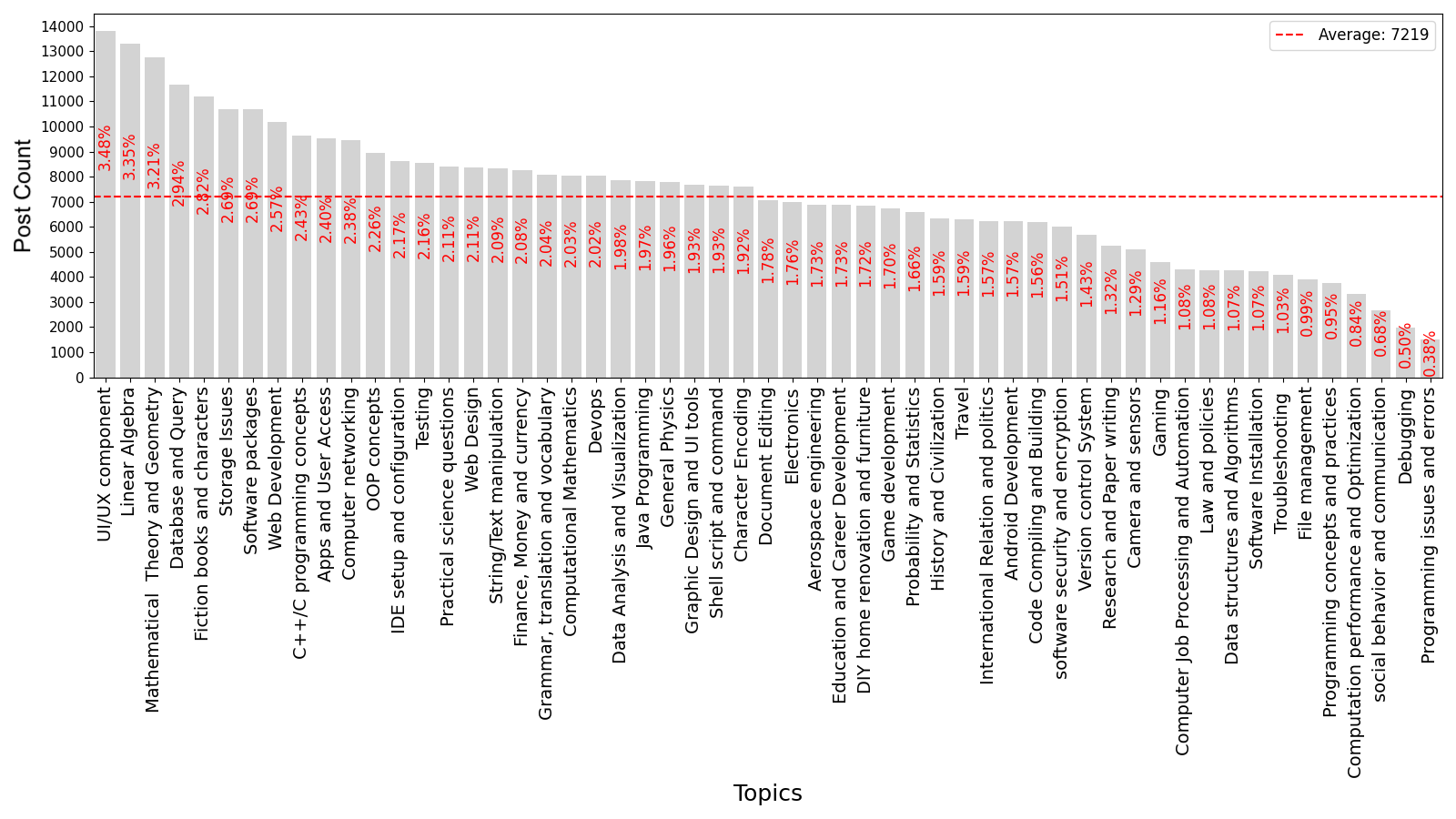}
    }
    \caption{Post Percentage and Topics }
    \label{fig:Post percentage and topics}
\end{figure*}
\subsection{Methodology}

%\oa{You didn't talk about how you collected the users. Also, we need a statistics table with data about number of users, posts, sites, answers, accepted answers etc. Also, you didn't mention that you only used accepted answers: Hi professor i added all the correction embeded in the description as table would take more space}

% \textbf{Figure~\ref{fig:Topic Modeling Overview Flow Chart}} illustrates the overview of the the differennt steps imvolved in our methodlogy. The key steps for generating topics from CS educators’ posts are outlined below: 1. Pre-process all posts (Data Preprocessing in \textbf{Figure~\ref{fig:Topic Modeling Overview Flow Chart}}), 2. Find the optimal number of topics (Create Dictionary and Corpus), 3. Generate the final topics (Topic Modeling and Manually Label Topics). 
Fig.~\ref{fig:Topic Modeling Overview Flow Chart} illustrates an overview of the different steps involved in our methodology. The key steps for generating topics from CS educators’ posts are outlined below: (1) pre-process all posts (Data Preprocessing in Fig.~\ref{fig:Topic Modeling Overview Flow Chart}), (2) determine the optimal number of topics (Create Dictionary and Corpus), and (3) generate the final topics (Topic Modeling and Manual Topic Labeling).

%We discuss these steps below.

\textbf{Step 1: Pre-process raw texts from CS educators' posts }

To enhance data quality, we followed standard pre-processing steps aligned with prior work \cite{abdellatif2020, bagherzadeh2019}. We wrote a function to remove non-text content such as code snippets and HTML tags (e.g., \texttt{<code>}, \texttt{<p>}, \texttt{<a>}) and then removed stop words, numbers, punctuation, and non-alphabetic characters using NLTK \cite{nltk2016}. We then applied spaCy-based lemmatization \cite{spacy_lemmatizer_api} to normalize tokens to their linguistically meaningful base forms. For instance, verb inflections such as “managed” and “managing” are reduced to “manage”. In contrast, the noun “management” remains unchanged, as spaCy performs context-aware lemmatization that preserves semantic meaning rather than mechanically stripping suffixes. A similar behavior is observed for irregular forms, where “running,” “ran,” and “runs” are all normalized to “run”, while nouns such as “runner” are retained. This linguistically informed normalization reduces vocabulary sparsity while maintaining semantic coherence, which ensures cleaner input for coherent topic modeling.

\textbf{Step 2: Find the Optimal Number of Topics}

We used Latent Dirichlet Allocation (LDA) via the MALLET library \cite{mccallum2002mallet} to generate topics from the pre-processed corpus. The model groups posts into \emph{K} topics, with the optimal \emph{K} selected using the coherence-based method by Arun et al. \cite{arun2010}. This approach ensures the resulting topics are distinct and interpretable.

Following prior research \cite{blei2003, rder2015}, we assessed topic quality using the \emph{c\_v} coherence metric to ensure semantic consistency. We ran the MALLET LDA model with \emph{K} values ranging from 5 to 70 (in steps of 5), computing coherence for each. The best score, 0.6318, occurred at \emph{K} = 55, which we chose as the optimal number of topics. Fig.~\ref{fig:Topics-Coherence score graph} shows the graph of the coherence scores for the different number of topics.

\textbf{Step 3: Generate Final Topics}

After determining the optimal number of topics, we ran the LDA model to extract 55 topics from the processed dataset. For each topic, we gathered: 
\begin{itemize}
  \item \textbf{Top Words}: The 50 terms with the highest probabilities representing the topic.
  \item \textbf{Relevant Posts}: Posts most closely aligned with the topic, scored between 0 and 1 by relevance.
\end{itemize}
This step helped organize 79,854,463 posts (including 32,187,805 questions and 47,666,658 answers), highlighting contributions from 12,949 users, i.e., CS educators. 

To interpret the extracted topics, we manually assigned descriptive labels based on their central themes. Using the open card sorting method \cite{hudson2013}, consistent with prior work \cite{abdellatif2020, agrawal2018, bagherzadeh2019, yang2016}, labels were derived iteratively from two sources: (1) the top associated words and (2) a random sample of 20--25 questions per topic. A student researcher and a faculty advisor collaboratively reviewed and refined the topic labels through over 20 Zoom-based iterations. In total, we finalized 55 topics and grouped them into categories.
%The topics spanned a wide range—from programming and software tools to mathematics, education, and humanities—capturing both IT and non-IT areas of interest.

\begin{figure*}[t]
    \centering
    \includegraphics[width=.85\textwidth]{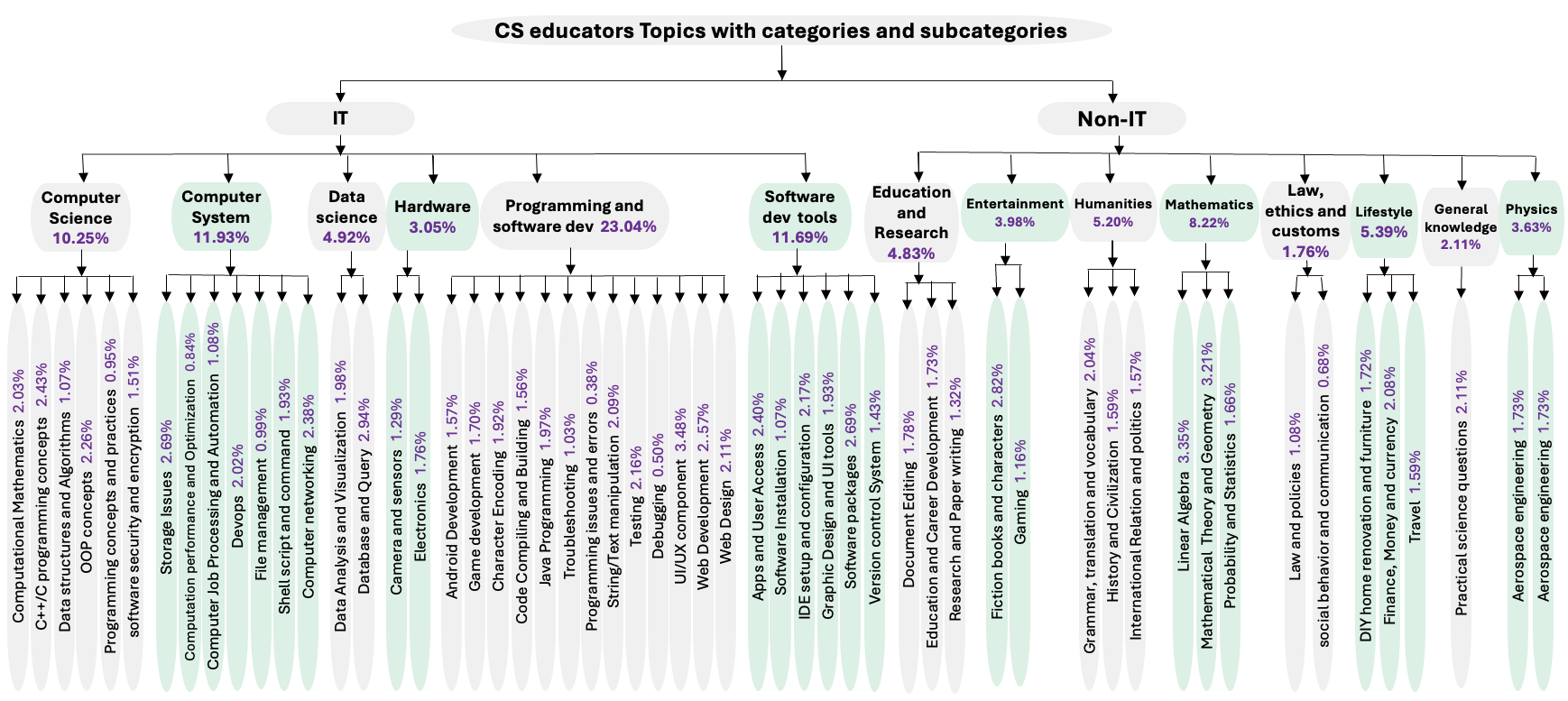}
    \caption{Hierarchical View of Topics, Categories and Subcategories}
    \label{fig:hierarchical_view_topics_categories}
\end{figure*}

\section{RQ1: What topics are discussed by CS educators across all sites of Stack Exchange network?}

 Fig. ~\ref{fig:Post percentage and topics} offers an overview of topic distributions, including average question counts per topic. After labeling, we grouped related topics into broader categories. For instance, "Web Development" and "Android Development" were placed under the \textbf{IT} category, along with other topics like "Testing" and "Code Compiling". These topics were grouped under Programming and Software Development subcategory. Likewise, in the \textbf{Non-IT} category, mathematical topics were divided into areas such as "Linear Algebra" and "Probability and Statistics".

This hierarchical classification continued until no further meaningful groupings emerged. Topics under "Computer Science," such as "Programming Concepts" and "Data Structures and Algorithms," were merged under general programming. The full structure is illustrated in Fig. ~\ref{fig:hierarchical_view_topics_categories}.

\subsection{IT Category}
The \textbf{IT} category accounted for 64.88\% of the posts, which were further divided into several subcategories, as discussed below.

%\begin{itemize}
 \textbf{Programming and Software Development (23.00\%)}: Topics covered include \textit{Game Development}, \textit{String/Text Manipulation}, \textit{Java Programming}, and \textit{Testing and Debugging}. Posts reflect varied development scenarios, ranging from language constructs to UI behavior and error handling. For example, one user discussed defining composite unique constraints using the SQL Server Management Studio GUI (\href{https://stackoverflow.com/questions/1670708}{$Q_{1670708}$}). %Another examined missing JSON payloads in ASP.NET MVC 5 for non-200 responses, suggesting adjustments in \texttt{web.config} (\href{https://stackoverflow.com/questions/45097867}{$Q_{45097867}$}).

   \textbf{Computer Systems (11.96\%)}: This subcategory  focuses on \textit{Computation Performance and Optimization}, \textit{DevOps}, and \textit{Computer Networking}. Users often encountered configuration issues arising from differences in system environments. One user found that Visual Studio Code failed to detect the \texttt{fish} shell when launched via an Automator script on macOS, due to inconsistent environment variable loading (\href{https://apple.stackexchange.com/questions/435021}{$Q_{435021}$}).
    
   \textbf{Computer Science (10.2\%)}: This subcategory includes topics like \textit{Programming Concepts and Practices}, \textit{Data Structures and Algorithms}. One user explored the limitations of using \texttt{operator ""} for user-defined literals in C++ within namespace declarations (\href{https://stackoverflow.com/questions/32319308}{$Q_{32319308}$}).

  \textbf{Software Development Tools (11.7\%)}: This subcategory includes topics on \textit{Version Control Systems}, \textit{IDE Setup and Configuration}, and \textit{Software Packages}. Posts often involve challenges with tools, debugging, and cross-environment behavior. One user struggled with \texttt{mousePressed()} in the p5.js tool due to incorrect context binding (\href{https://stackoverflow.com/questions/51222639}{$Q_{51222639}$}). 

    \textbf{Data Science (4.92\%)}: This subcategory contains two topics: \textit{Data Analysis and Visualization} and \textit{Database and Query Systems}. They often discuss issues related to performance and data modeling. One user migrating from MS Access explored SQL Server features like multi-statement table-valued functions and indexed views to improve reporting efficiency (\href{https://stackoverflow.com/questions/6998635}{$Q_{6998635}$}).

     \textbf{Hardware (3.0\%)}: Hardware-related topics cover troubleshooting, embedded electronics, and cross-platform interfacing. One Lenovo user investigated the effect of shorting the data and +20V pins in a charging port, aiming to understand the adapter’s data pin role (\href{https://superuser.com/questions/1537370}{$Q_{1537370}$}).

%\end{itemize}

\subsection{Non-IT Category}
The \textbf{Non-IT} category represented 35.12\% of the posts as shown in Fig. 3. The subcategories within this category are the following:
%\begin{itemize}
    
    \textbf{Education and Research (4.83\%)}: This subcategory includes topics on \textit{Document Editing}, \textit{Education and Career Development}, and \textit{Research and Paper Writing}. One example involved resolving citation issues in a \texttt{psmatrix} environment in \LaTeX, which was fixed by enabling the \texttt{runs=2} option in \texttt{auto-pst-pdf} for proper label resolution during compilation (\href{https://tex.stackexchange.com/questions/266900}{$Q_{266900}$}).

     \textbf{Entertainment (3.98\%)}: This subcategory covers game mechanics, trivia, and creative programming puzzles. For example, here is a user asking about poker rules (\href{https://poker.stackexchange.com/questions/10645}{$Q_{10645}$}).
    
     \textbf{General knowledge (2.1\%)}: This subcategory involves everyday scientific inquiries and practical problem solving. One user questioned the biological reasoning behind food aversions, asking if unpleasant taste could signal incompatibility or potential harm for some individuals (\href{https://biology.stackexchange.com/questions/51819}{$Q_{51819}$}).

     \textbf{Mathematics (8.22\%)}: This sub-category includes topics like \textit{Mathematical Theory and Geometry}, \textit{Probability and Statistics}, and \textit{Linear Algebra}. One user discussed the difference between a Galois extension being solvable by radicals versus being a radical extension (\href{https://math.stackexchange.com/questions/598285}{$Q_{598285}$}).

   \textbf{Humanities (5.20\%)}: This subcategory includes topics like \textit{History,  Civilization} and \textit{International Relations}. A user questioned whether the U.S. Electoral College contradicts the Constitution (\href{https://politics.stackexchange.com/questions/3296}{$Q_{3296}$}).
   
   %, especially in light of voting equality and the 9th Amendment. The response clarified that the Electoral College is explicitly established in Article II, and while some argue it creates unequal representation, courts have generally upheld its constitutionality (\href{https://politics.stackexchange.com/questions/3296}{$Q_{3296}$}).
   
 \textbf{Law, ethics and customs (1.8\%)}: This subcategory contains two topics: law and policies, and social behavior and communication. One user analyzed the meaning of “I’m gonna serve it to you” interpreting it as a form of verbal retaliation grounded in idiomatic and legal usage of “serve” (\href{https://english.stackexchange.com/questions/156045}{$Q_{156045}$}).

  \textbf{Life Style (5.39\%)}: Encompasses topics like "Finance," "Travel," and "DIY home improvement." One user asked about prepaid SIM card use across different countries (\href{https://travel.stackexchange.com/questions/20594}{$Q_{20594}$}).

    % In your document, where you want the figure

     \textbf{Physics (3.63\%)}: This subcategory encompasses both experimental and theoretical questions in classical and modern physics. A user sought to understand the cause of light’s apparent slowing in transparent media (\href{https://physics.stackexchange.com/questions/130567}{$Q_{130567}$}). 
     
     %Another asked about the alignment of planetary orbits and galactic structure, inquiring whether stars and planets rotate in planes co-aligned with the Milky Way disk (\href{https://astronomy.stackexchange.com/questions/40755}{$Q_{40755}$}).

%\end{itemize}

\paragraph{Summary of RQ1.}
CS educators discuss a variety of topics across the Stack Exchange network, ranging from programming discussions to lifestyle questions. Our analysis identified 55 unique topics, grouped into IT (64.88\%) and Non-IT (35.12\%) categories. Within IT, the largest subcategory was Programming and Software Development (23.04\%), featuring topics like "Web Development" and "Game Development." In Non-IT, Mathematics (8.22\%) was most prominent, including topics such as "Linear Algebra" and "Probability and Statistics". Other Non-IT topics include Education and Research, and Law and Ethics.

% \textbf{Summary of RQ1: What topics are discussed by CS educators across Stack Exchange sites?}

% Our analysis identified \textbf{55 unique topics}, grouped into \textbf{IT (64.88\%)} and \textbf{Non-IT (35.12\%)} categories. Within IT, the largest subcategory was \textbf{Programming and Software Development (23.04\%)}, featuring topics like "Web Development" and "Game Development." In Non-IT, \textbf{Mathematics (8.22\%)} was most prominent, including areas such as "Linear Algebra" and "Probability and Statistics."

\section{RQ2: How do topics discussed by CS educators evolve over time?}

%\oa{Aren't these steps should belong to Section 3?: not it belongs to this section, Make sure that these are not copied from the IoT paper. You need to re-write them: rewrote the formulas, added a reference }
We assessed each topic’s importance using two metrics: absolute and relative impact. Absolute impact measures the number of new posts added each month. Relative impact refers to the number of new posts added to a category each month relative to other categories. We used these metrics as applied in other studies on Stack Overflow and Stack Exchange~\cite{uddin2021empirical, abdellatif2020, bagherzadeh2019}.

\textbf{Absolute Impact for a category:}
We apply Latent Dirichlet Allocation (LDA) to the corpus \( a_j \) to generate a set of \( N \) topics \( (x_1, \dots, x_n) \). For a specific topic \( x_n \) in a post \( p_i \), the number of posts related to topic \( x_n \) is denoted by \( I(p_i, x_n) \), where \( I(p_i, x_n) \) is an indicator function that returns 1 if topic \( z_n \) is the dominant topic in post \( p_i \), and 0 otherwise.

The absolute impact metric for a topic \( x_n \) in a month \( m \) is computed using the following formula:

\begin{equation}
    \text{impact}_{\text{absolute}}(x_n, m) = \sum_{i=1}^{P(m)} I(p_i, x_n)
\end{equation}

where:
 \( P(m) \) is the total number of posts in month \( m \),
 \( I(p_i, x_n) \) is the indicator function that returns 1 if topic \( x_n \) is the dominant topic in post \( p_i \), and 0 otherwise.

The indicator function \( I(p_i, x_n) \) is defined as:

\begin{equation}
    I(p_i, x_n) =
    \begin{cases}
        1 & \text{if } x_n \text{ is the dominant topic in post } p_i, \\
        0 & \text{otherwise.}
    \end{cases}
\end{equation}

%Now lets Compute Absolute Impact for a two main category.

To compute the absolute impact for a category \( C \), which contains multiple topics, we sum the absolute impacts for all topics \( x_n \) within that category. The formula for this is:

\begin{equation}
    \text{impact}_{\text{absolute}}(C, m) = \sum_{x_n \in C} \text{impact}_{\text{absolute}}(x_n, m)
\end{equation}

where:
 \( C \) is a category (such as "IT" or "Non-IT"),
 \( x_n \in C \) denotes all topics \(x_n \) that belong to the category \( C \).

\textbf{Relative Impact for a Category}

To calculate the relative impact of a specific topic over time, we use a relative impact metric. The relative impact metric of a topic \( x_n \) in month \( m \) is defined as follows:

\begin{equation}
    \text{impact}_{\text{relative}}(x_n, m) = \frac{1}{P(m)} \sum_{i=1}^{P(m)} I(p_i, x_n)
\end{equation}

where \( P(m) \) is the total number of posts in month \( m \) that contain the topic \( x_n \), and \( I(p_i, x_n) \) is the indicator function that equals 1 if the topic \( x_n \) is present in post \( p_i \), and 0 otherwise.

We also compute the relative impact for a category \( C \) in month \( m \) by summing the relative impacts of all topics \( x_n \) within that category:

\begin{equation}
    \text{impact}_{\text{relative}}(C, m) = \sum_{x_n \in C} \text{impact}_{\text{relative}}(x_n, m)
\end{equation}

Here, \( C \) is one of the two major categories IT and Non-IT.

\subsection{Results:}

Using the previous equations, we calculated the absolute and relative impact of the specific categories discussed by CS educators between March 2008 and May 2024, based on our dataset. This subsection explores the trends for the IT and Non-IT categories over time.

\subsubsection{ \textbf{Absolute Impact}}

\begin{figure}
    \centering
    \includegraphics[width=1\linewidth]{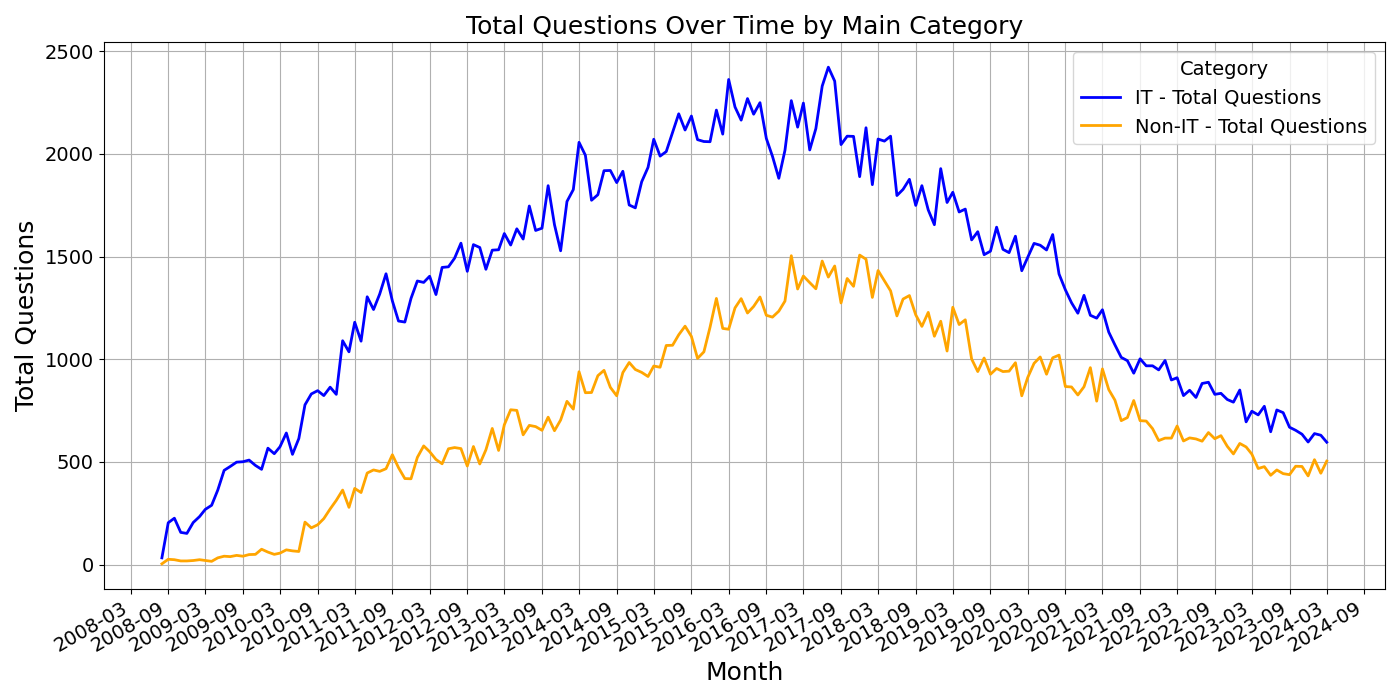}
    \caption{Absolute Impact of Main Categories}
    \label{fig:abs impact main categories}
\end{figure}

We examined the absolute impact of topics using the approach outlined in Equation 3. As shown in Fig~\ref{fig:abs impact main categories}, the number of posts over time highlights that the IT category has consistently led the way from 2008 to 2024. There were clear spikes in IT-related activity around 2014 and again in 2017. While the Non-IT category had fewer overall discussions, it showed steady growth—especially between 2013 and 2018. Within the IT category, Programming and Software Development had the strongest presence, with steady growth from 2009 and significant peaks in 2014 and 2017.

Other subcategories, such as Computer Systems and Software Development Tools, also saw consistent engagement over the years. Interestingly, Hardware saw its highest activity around 2016, possibly reflecting increased interest in computing hardware among educators. In the Non-IT space, Mathematics consistently received the most attention, followed by Lifestyle, Humanities and Education and Research. The figures for the subcategories are not shown due to space limitations.

%These patterns suggest a broader trend toward integrating general education subjects with computing, reflecting a growing interdisciplinary approach in CS education.

% We examined the absolute impact of topics using the approach outlined in Equation 3. As shown in Figure~\ref{fig:abs impact main categories}, the number of posts posted over time highlights how posts in the IT category have consistently led the way from 2008 to 2024. There were clear spikes in IT-related activity around 2014 and again in 2017. While the Non-IT category had fewer overall discussions, it showed steady growth—especially between 2013 and 2018. Within the IT category, Programming and Software Development had the strongest presence, with steady growth from 2009 and significant peaks in 2014 and 2017.  

% Other subcategories, such as Computer Systems and Software Development Tools, also saw consistent engagement over the years. Interestingly, Hardware saw its highest activity around 2016, possibly reflecting increased interest in computing hardware by educators. In the Non-IT space, Mathematics consistently received the most attention, followed by Humanities and Education and Research. These patterns suggest a broader trend toward integrating general education subjects with computing, reflecting a growing interdisciplinary approach in CS education.

\subsubsection{Relative Impact}

\begin{figure}
    \centering
    \includegraphics[width=\linewidth]{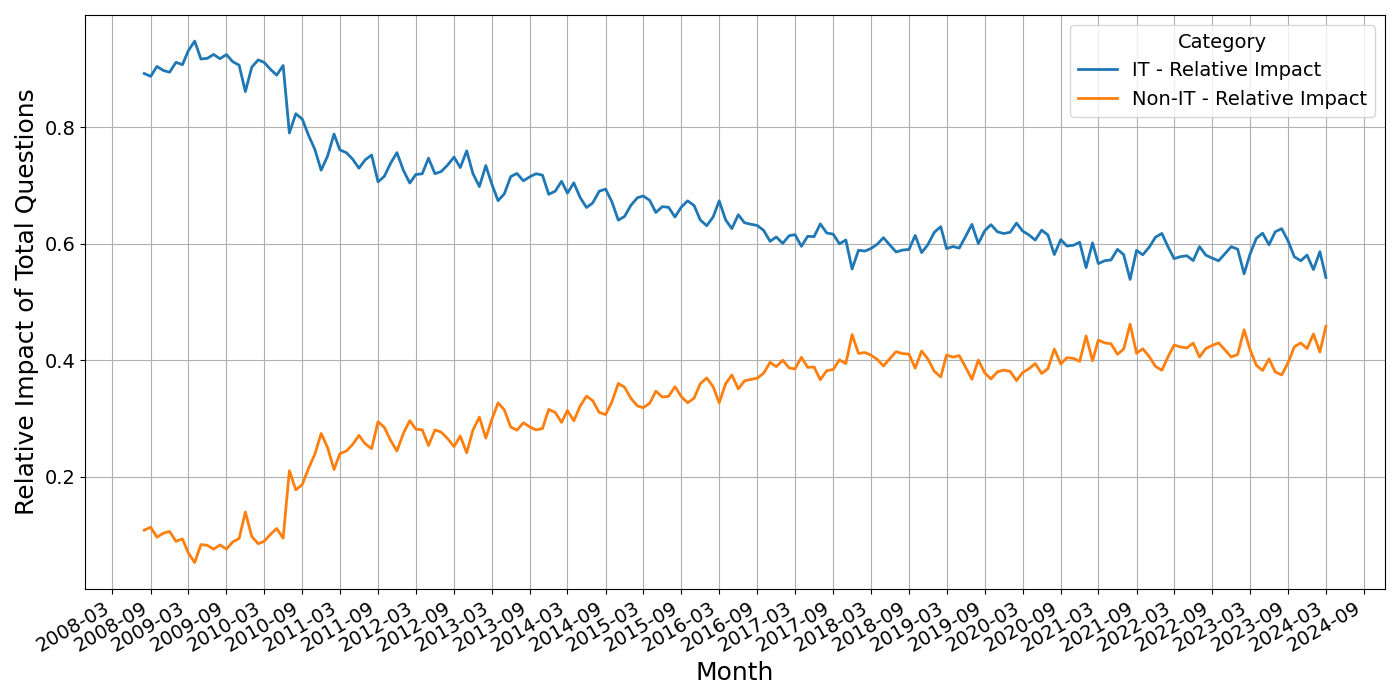}
    \caption{Relative Impact of Main Categories}
    \label{fig:relative impact main cat}
\end{figure}

We calculated the relative impact of different topic categories using the method described in Equation 5. As illustrated in Figure~\ref{fig:relative impact main cat}, IT topics led the conversation between 2009 and 2012. However, from 2014 onward, Non-IT topics began to gain ground in relative popularity. This shift highlights a growing interest in educational and interdisciplinary discussions alongside traditional technical content.  

Among the IT subcategories, Programming and Software Development consistently stood out with the highest levels of engagement. Around 2014, we began to see growing interest in areas like Software Development Tools and Data Science. On the Non-IT side, Mathematics attracted steady attention, with peaks in 2017 and 2018. There’s also been a noticeable increase in discussions around Humanities and Education and Research. We plan to examine trends in subcategories and topics in greater detail in future work.

%, pointing to a broader push to integrate these fields into computer science education.

%Overall, our findings highlight meaningful trends not only at the category level but also within specific subcategories, revealing how educator engagement varies across different content areas and how priorities have evolved over time.

\paragraph{Summary of RQ2:}
IT-related topics dominate the discussions by CS educators, especially those related to programming. However, Non-IT categories experienced steady growth from 2013 to 2018. Non-IT topics, particularly those related to Mathematics, Lifestyle, and Education and Research, gained increased attention.

\section{Threats to Validity}

\textbf{External Validity:} 
Threats related to the generalizability of our findings. Although the CS Educators community comprises a large number of educators, our results may not generalize to all CS educators. Furthermore, our analysis focuses on CS educators' discussions on Stack Exchange, a very large discussion forum. However, our findings may not generalize to other online discussion forums in which CS educators engage.

%The generalizability of our findings is limited by the specific context of this study. We focus on Workplace StackExchange, one of the largest and most popular Q\&A websites for workplace discussions among developers. Our findings may not generalize to developers who are not active in online forums. 

%Previous studies~\cite{xie2020understanding}, have noted that Q\&A platforms in other domains may exhibit distinct user behavior patterns and content dynamics. These differences may affect how questions are posed and answered in those contexts. Therefore, further research is needed to determine whether similar trends exist among developers who are not involved in Q\&A platforms.

\textbf{ Internal Validity: } 
Our study involves manual labeling and sorting into categories, which may introduce biases and errors. However, we aimed to minimize these biases by using two raters. 

 \textbf{Construct Validity:} We collected a very large dataset of posts contributed by users on the CS Educators site. However, some users may only read posts without actively posting or participating. One could argue that these users are still relevant, regardless of whether they contribute content, as their presence reflects interest in CS education topics. Further analysis of user participation and activity patterns is needed, as users may exhibit different engagement behaviors across the Stack Exchange network~\cite{xie2020understanding}.

Another limitation of our study is that the dataset extends only through 2024. Consequently, data from 2025 are not included. Although our analysis spans over 16 years of data—suggesting that the identified topics are relatively stable—the absence of 2025 data means that the results may not fully reflect the most recent trends.

 %We collected a very large dataset of posts by users on the CS Educators site. However, there may be users who only read posts without actively posting or participating. One could argue that these users are still relevant, regardless of whether they contribute, since they are interested in CS education topics. Further analysis is needed on user participation and activity patterns, as users may exhibit different engagement behaviors across Stack Exchange~\cite{xie2020understanding}. Another limitation of our study is that our dataset is dated until 2024, therefore, we are missing the 2025 data. One could argue that our analysies spanned over 16 years of data which makes the topics identified stable, nevertheless, the 2025 missing data makkes the result not very recent.

%Threats related to the difficulty in finding data relevant to workplace discussions are notable. We collected all posts from the Workplace StackExchange site, avoiding the use of tags, which previous research relied on~\cite{uddin2021,tahir2020}, as our dataset from the site is homogeneous. 

\section{Future Works and Conclusion}

This study offers a preliminary analysis into how CS educators participate across the Stack Exchange network. We collected posts written by users of the CS Educators site across all Stack Exchange platforms. We conducted topic modeling on a dataset of over 79 million posts by approximately 13,000 CS educators. We manually labeled the topics and categorized them into different categories and subcategories. We found that IT-related topics dominate the discussions, while non-IT topics showed steady growth between 2013 and 2018.

In the future, we plan to conduct an in-depth analysis of topic trends at the subcategory and topic levels. We also plan to carry out a detailed analysis of user participation, examine the most active contributors, and perform focused studies on topics discussed by CS educators on targeted sites, such as the CS Educators site. Additionally, we plan to investigate the relative difficulty and popularity of different topics and subcategories.

\bibliographystyle{ACM-Reference-Format}
\bibliography{bibliography}

\end{document}